\documentclass[
  aps, prb,
  twoside,
  twocolumn,
  showpacs,
  floatfix,
  superscriptaddress,
  10pt,
  citeautoscript,
]{revtex4-1}

\usepackage{graphicx}
\usepackage{amsmath}
\usepackage{color}
\usepackage{times}

\newcommand{\mat}[1]{\ensuremath{\mathbf{#1}}}
\newcommand{\eV}{\text{eV}}
\renewcommand{\deg}{\ensuremath{^\circ}}

\newcommand{\W}{\text{W}}

\newcommand{\sect}[1]{Sect.~\ref{#1}}
\newcommand{\fig}[1]{Fig.~\ref{#1}}

\newcommand{\eq}[1]{Eq.~(\ref{#1})}

\newcommand{\myscale}{0.62}

\begin{document}

\author{Leili Gharaee}
\affiliation{
  Chalmers University of Technology,
  Department of Physics,
  S-412 96 Gothenburg, Sweden
}
\author{Jaime Marian}
\affiliation{
  University of California,
  Department of Materials Science and Engineering,
  Los Angeles, California 90095, United States of America
}
\author{Paul Erhart}
\email{erhart@chalmers.se}
\affiliation{
  Chalmers University of Technology,
  Department of Physics,
  S-412 96 Gothenburg, Sweden
}

\title{
  The role of interstitial binding in radiation induced segregation in W-Re alloys
}

\begin{abstract}
Due to their high strength and advantageous high-temperature properties, tungsten-based alloys are being considered as plasma-facing candidate materials in fusion devices. Under neutron irradiation, rhenium, which is produced by nuclear transmutation, has been found to precipitate in elongated precipitates forming thermodynamic intermetallic phases at concentrations well below the solubility limit. Recent measurements have shown that Re precipitation can lead to substantial hardening, which may have a detrimental effect on the fracture toughness of W alloys. This puzzle of sub-solubility precipitation points to the role played by irradiation induced defects, specifically mixed solute-W interstitials. Here, using first-principles calculations based on density functional theory, we study the energetics of mixed interstitial defects in W-Re, W-V, and W-Ti alloys, as well as the heat of mixing for each substitutional solute. We find that mixed interstitials in all systems are strongly attracted to each other with binding energies of $-2.4$ to $-3.2\,\eV$ and form interstitial pairs that are aligned along parallel first-neighbor $\left<111\right>$ strings. Low barriers for defect translation and rotation enable defect agglomeration and alignment even at moderate temperatures. We propose that these elongated agglomerates of mixed-interstitials may act as precursors for the formation of needle-shaped intermetallic precipitates. This interstitial-based mechanism is not limited to radiation induced segregation and precipitation in W--Re alloys but is also applicable to other body-centered cubic alloys.
\end{abstract}

\date{\today}


\maketitle

\section{Motivation}

Tungsten is being considered as a candidate material in magnetic fusion energy devices due to its high strength and excellent high temperature properties \cite{ZinGho00, RieDudGon13, FitNgu08, BecDom09}. Upon fast neutron irradiation in the 600 to 1000$^{\circ}$ C temperature range, W transmutes into Re by way of beta decay reactions at a rate that depends on the neutron spectrum and  the position in the reactor. For the DEMO (demonstration fusion power plant) reactor concept, calculations show that the transmutation rate is 2000 and 7000 atomic parts per million (appm) per displacements per atom (dpa) in the divertor and the equatorial plane of the first wall, respectively (where damage, in each case, accumulates at rates of 3.4 and 4.4 dpa/year) \cite{GilSub11, GilDudZhe12}.

The irradiated microstructure initially evolves by accumulating a high density of prismatic dislocation loops and vacancy clusters, approximately up to 0.15 dpa \cite{TanHasHe09, HasTanNog11}. Subsequently, a void lattice emerges and fully develops at fluences of around 1 dpa. After a critical dose that ranges between 5 dpa for fast ($>$1 MeV) neutron irradiation \cite{HasTanNog11} and 2.2 dpa in modified target rabbits in the HFIR \cite{HuKoyKat15}, W and W-Re alloys develop a high density of nanometric precipitates with acicular shape at Re concentrations well below the solubility limit \cite{HasTanNog11}. The structure of these precipitates is consistent with $\sigma$ (W$_7$Re$_6$) and $\chi$ (WRe$_3$) intermetallic phases, which under equilibrium conditions only occur at temperatures and Re concentrations substantially higher than those found in neutron irradiation studies. A principal signature of the formation of these intermetallic structures in body-centered cubic (BCC) W is the sharp increases in hardness and embrittlement, as well as other detrimental effects \cite{HasTanNog11, TanHasHe07}. Qualitatively similar observations were recently made on W-25Re alloys subjected to heavy ion irradiation \cite{XuBecArm15}, clearly establishing a link between primary damage production and Re precipitation. 

\begin{figure}
  \centering
  \includegraphics[scale=0.15]{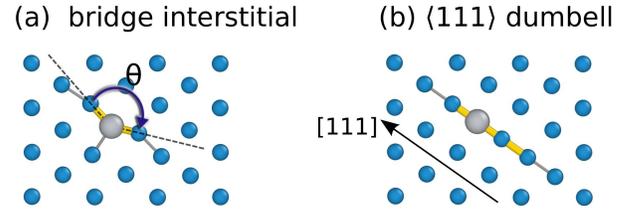}
  \caption{
    Bridge (a) and $\left<111\right>$ dumbbell (b) interstitial defects in tungsten. The $\left<111\right>$ crowdion configuration closely resembles the $\left<111\right>$ dumbbell configuration with a slightly larger spacing of the defect atoms along the $\left<111\right>$ axis. The figure shows a slice parallel to $\left\{110\right\}$. Small (blue) spheres indicate tungsten atoms whereas large (gray) spheres indicate extrinsic atoms (V, Ti, Re). Thicker (yellow) cylinders indicate bond lengths shorter than 2.3\,\AA\ whereas thinner (gray) cylinders indicate bond lengths shorter than 2.5\,\AA. The bond angle $\theta$ is indicated in (a).
  }
  \label{fig:single-interstitials}
\end{figure}

For a number of reasons, the formation of these non-thermodynamic phases is inconsistent with the standard picture of solute precipitation by a vacancy mechanism, even in the context of radiation enhanced diffusion (RED) and radiation induced precipitation (RIP). For example, the phase diagram points to the $\sigma$ phase as the first one to emerge from the BCC solid solution, whereas electron diffraction studies show the co-occurrence of both phases, with even a initial preeminence of the $\chi$ phase in some cases \cite{WilWifBen83}. All this is suggestive of the role played by interstitial defects, such as mixed dumbbells and crowdions, in facilitating solute transport and potentially shifting phase boundary lines. Although detailed kinetic models linking solute transport and precipitation with irradiation defects exist (e.g., Ref.~\onlinecite{Ard08}), self- and mixed-interstitial (solute) transport introduces a set of particularities that cannot be captured in effective models. Specifically, motion directionality and solute transport mechanisms are directly governed by crystal structure and the chemical nature of the defect-solute binding. These are processes that must be characterized at the atomistic scale using first principles calculations. There have been recent efforts to study interstitial defect energetics in W-based alloys, using density functional theory (DFT) calculations, in an attempt to establish the grounds for defect diffusion models \cite{BecDom09, SuzYamHas15, GhaErh15}. These studies build on existing knowledge of self-interstitial atoms (SIA) in metals gained from electronic structure and semi-empirical calculations over the last 20 or so years \cite{WilFuMar05,DerNguDud07}.

In any case, the formation of sub-soluble intermetallic phases in irradiated W and W-Re remains unexplained, and, to date, a detailed model of Re precipitation in irradiated W is still lacking. In this paper, we study the precipitation behavior of W--V, W--Ti and W--Re alloys as a function of alloy composition with an emphasis on interstitials in order to understand and explain property changes of tungsten under irradiation conditions. The paper is organized as follows. In the next section we describe the general approach as well as the computational methodology employed in this work.  The results of interstitial binding in the dilute limit for V, Ti and Re are given in \sect{sect:results-interstitial-binding}. This is followed by a comprehensive investigation of the mixing energies of both substitutional and interstitial-based systems over the full composition range in \sect{sect:results-mixing-energies} before we then consider in \sect{sect:results-migration} the mobility of mixed interstitials. Finally, we formulate a mechanism that provides a rationale for the occurrence and shape of experimentally observed $\sigma$ and $\chi$-phase precipitates in W--Re alloys.

\section{Methodology}
\label{sect:computational-methods}

\subsection{General approach}
\label{sect:method-structure-generation}

The present work is primarily centered around Re since as described above W--Re alloys are naturally formed due to nuclear transmutation under fusion conditions and are well known to exhibit RIP and RIS (radiation induced segregation) \cite{SuzYamHas14}. Vanadium and titanium are included as well since it was determined in Ref.~\onlinecite{GhaErh15} that their mixed interstitial configurations exhibit a number of similarities with Re. Specifically, it was shown that all three elements adopt a bridge-like interstitial configuration [\fig{fig:single-interstitials}(a)] in the W matrix. The latter can be thought of as a dumbbell interstitial oriented along $\left<111\right>$ [\fig{fig:single-interstitials}(b)], in which the extrinsic atom is displaced along one of three $\left<211\right>$ directions perpendicular to the dumbbell orientation. This results in a bond angle with the nearest neighbors of approximately 150\deg\ as opposed to 180\deg\ in the case of the straight dumbbell interstitial [see Fig. 1(a)].
In addition, Re, Ti, and V trap SIAs, i.e. the reaction $X_\W + (\W-\W)_\text{int} \rightarrow (X-\W)_\text{int}$ is exothermic, which strongly affects the migration behavior of SIAs. Finally, for all three elements the defect formation volume tensor of the mixed interstitials is strongly anisotropic, which thus also applies to the strain field associated with these defects and provides a strong driving force for defect alignment.
To investigate this behavior further, we considered several different types of configurations, the construction of which is described in the following:

\paragraph*{Double-interstitial configurations.}
To quantify in\-ter\-sti\-tial-in\-ter\-sti\-tial interactions, defect configurations including two interstitials were created based on $4\times4\times4$ conventional supercells (128 atoms). Configurations were constructed by systematically varying the distance as well as the relative orientation (rotation) of two bridge interstitials, yielding more than 100 initially distinct configurations.

\paragraph*{Dilute substitution.}
The direct interaction of two Ti, V, or Re atoms in a body-centered cubic W matrix was studied using $4\times4\times4$ conventional supercells. Their relative separation was varied to extract their (pair-wise) interaction as a function of distance.

\paragraph*{Substitutional alloys.}
The energetics of concentrated alloys was determined in two ways. Firstly, $3\times3\times3$ conventional supercells (54-atoms) were employed, in which a number of extrinsic substitutional atoms was inserted corresponding to concentrations covering the entire concentration range in the case of V and the concentration range up to 50\%\ in the case of Ti and Re. The latter restriction was imposed since the BCC lattice structure is mechanically unstable for these elements, whence Ti and Re-rich supercells do not maintain the BCC structure upon relaxation. Secondly, in the case of W--Ti an alloy cluster expansion was constructed using the atomic alloy toolkit, from which a series of low energy structures was derived.

\paragraph*{Interstitial based structures.}
As discussed in detail below, the defect supercell calculations indicated strong interstitial-interstitial interactions suggestive of defect clustering. To explore this effect further additional structures were constructed in the following fashion \cite{ErhSadCar08}. First, ideal supercells were generated based on the primitive cell by applying an integer transformation matrix $\mat{P}$ that relates the primitive cell metric $\mat{h_p}$ to the supercell metric $\mat{h}$ according to $\mat{h} = \mat{P}\mat{h_p}$. Subsequently, all distinct configurations were constructed that result from insertion of a bridge interstitial on one lattice site when taking into account the orientation of the interstitial. In this fashion, 166 ``interstitial based'' structures were obtained.

\subsection{Computational details}
\label{sect:method-computational-details}

For all configurations described in the previous section, density functional theory (DFT) calculations were carried out using the projector augmented wave (PAW) method \cite{Blo94, KreJou99} as implemented in the Vienna ab-initio simulation package \cite{KreHaf93, KreHaf94, KreFur96a, KreFur96b}. Since interstitial configurations involve short interatomic distances ``hard'' PAW setups that include semi-core electron states were employed with plane wave energy cutoffs of 343 eV, 290 eV and 295 eV for V, Ti and Re respectively.

Exchange and correlation effects were described using the generalized gradient approximation \cite{PerBurErn96} while the occupation of electronic states was performed using the first order Methfessel-Paxton scheme with a smearing width of 0.2\,eV. The Brillouin zone was sampled using Monkhorst-Pack grids of at least $4\times 4 \times 4$. (A detailed discussion of the effect of different computational parameters on the results can be found in Ref.~\onlinecite{GhaErh15}). All structures were optimized allowing full relaxation of both ionic positions and cell shape with forces converged to below 15 meV/\AA. Migration barriers were computed using $4\times4\times4$ supercells and the climbing image-nudged elastic band method with up to five images \cite{HenUbeJon00}.

\section{Results}
\label{sect:results}

\subsection{Interstitial-interstitial binding}
\label{sect:results-interstitial-binding}

The SIA in tungsten has been shown to adopt a so-called crowdion configuration in which the atoms that form the defect are delocalized along $\left<111\right>$ directions of the lattice. Some extrinsic elements such as V, Ti and Re bind to SIAs and cause the interstitial to localize \cite{GhaErh15}. These mixed-interstitial defects have very large defect formation volumes that range from 1.2 to 1.6 times the volume per atom of the ideal structure. The associated strain field is oriented along $\left<111\right>$ and highly anisotropic, which is evident from the ratio of the largest and smallest eigenvalues of the formation volume tensor.

When the concentration of solute atoms increases, the strain field produced by individual bridge mixed--interstitials remains unchanged. However, the reduction in the average distance between solute atoms implies that mixed-interstitials interactions become important, which due to the anisotropy of the strain field can be expected to exhibit a strong directionality. The system can thus reduce its strain energy by rearranging both defect location and orientation.

To test this possibility, we first consider the interaction of pairs of mixed-interstitials of V, Ti and Re, which were constructed as described in \sect{sect:method-structure-generation}. Specifically, we calculate the binding energy, which is defined as the difference between the formation energy of the di-interstitial configuration $E^f(2[X-\W]_\W)$ and the sum of the formation energy of two individual mixed--interstitial defects, $2E^f([X-\W]_\W)$,
\begin{align}
  E_{2[X]}^b = E^f(2[X-\W]_\W) - 2E^f([X-\W]_\W).
  \label{eq:binding-energy}
\end{align}
Note that by this definition two interstitials are attracted to each other if the binding energy is negative.

\begin{figure}
  \centering
  \includegraphics[scale=\myscale,trim=0 38 0 0,clip]{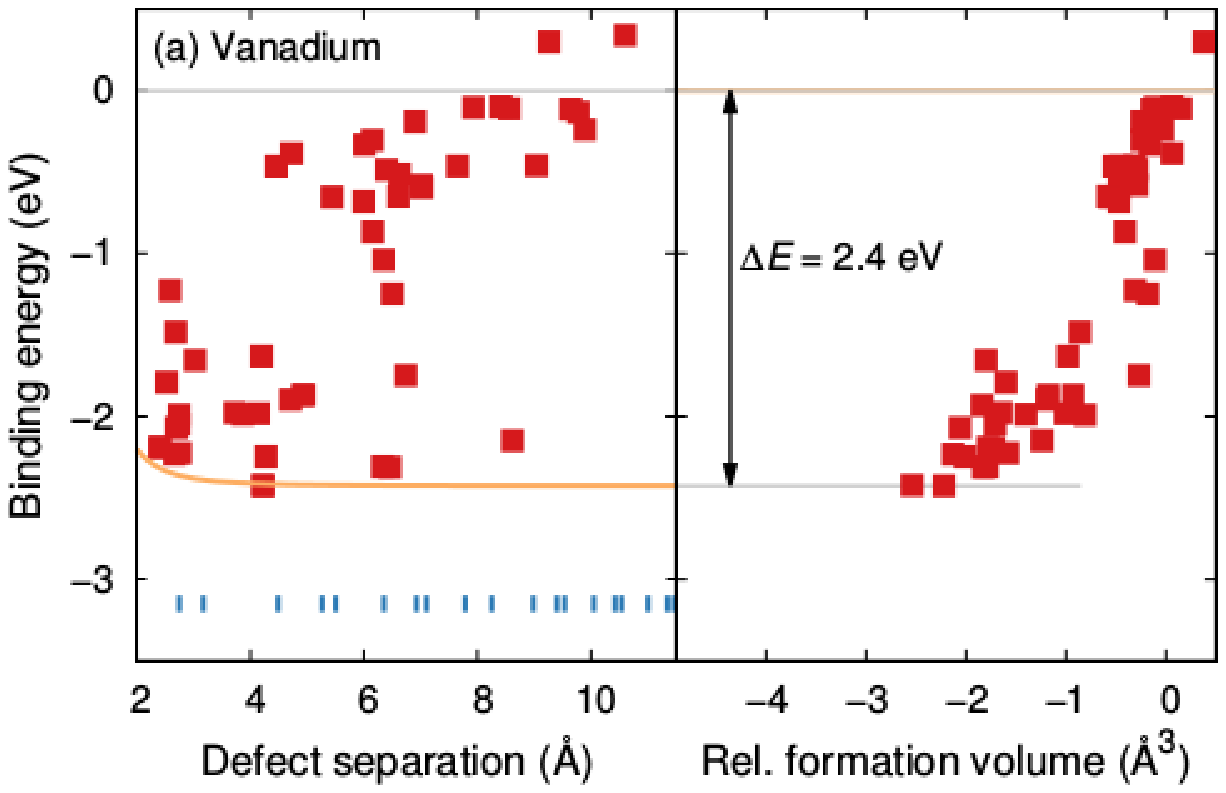}
  \includegraphics[scale=\myscale,trim=0 38 0 0,clip]{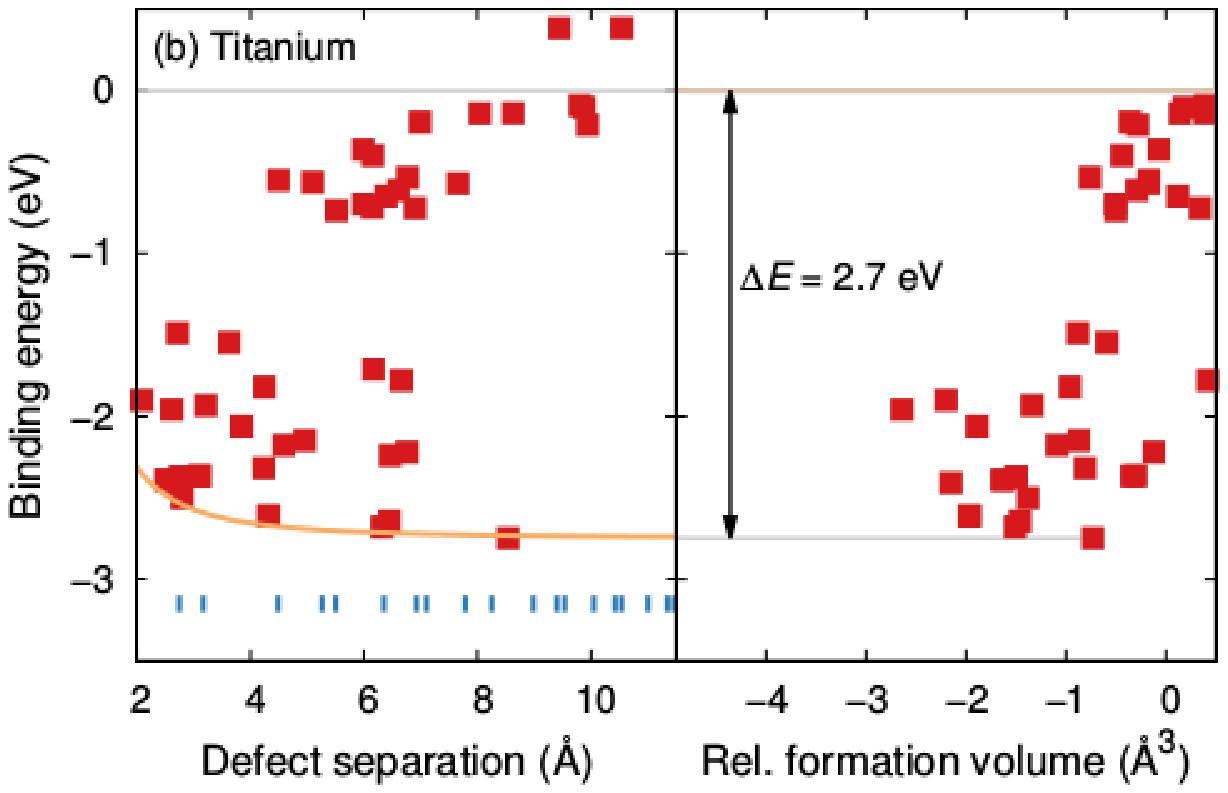}
  \includegraphics[scale=\myscale]{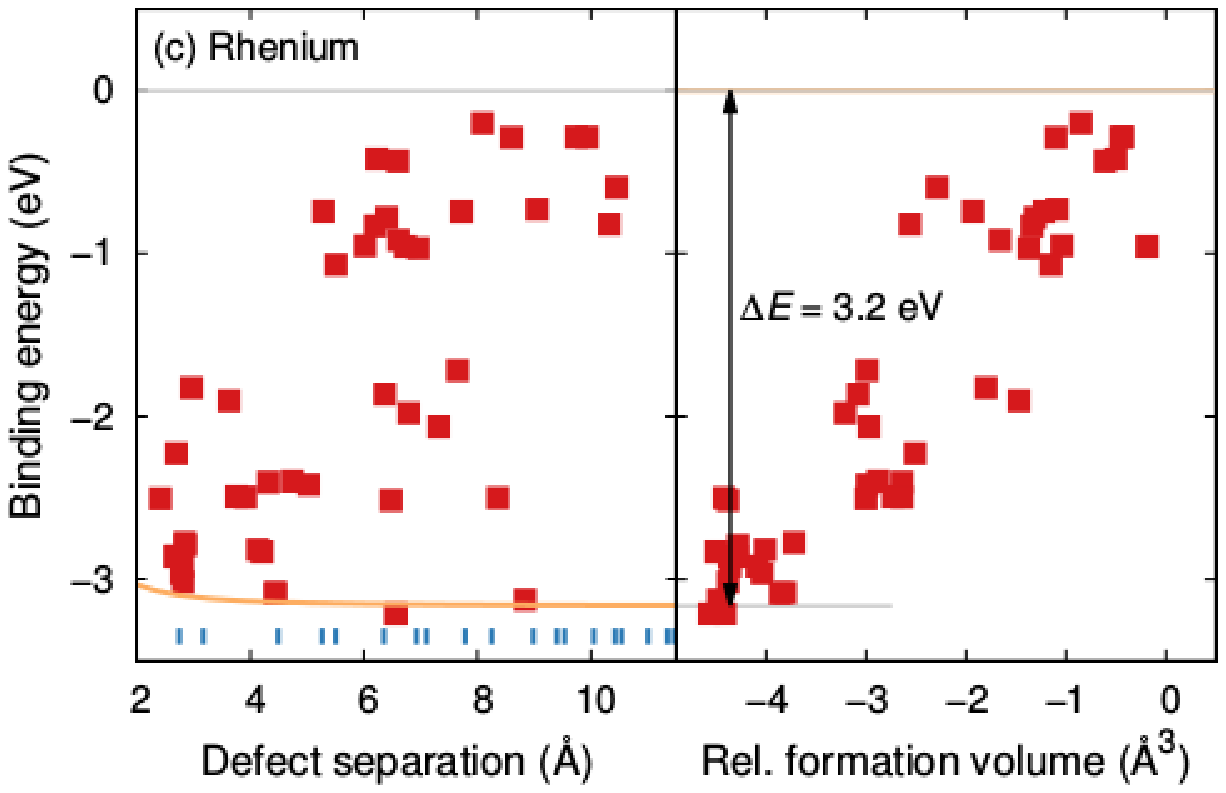}
  \caption{
    Binding energy according to \eq{eq:binding-energy} between two (a) V--W, (b) Ti--W, and (c) Re--W mixed interstitials as a function of pair separation (left) as well as relative formation volume (right), where latter was computed according to \eq{eq:formation-volume}. The blue tics indicate the positions of the neighbors shells in the perfect structure. The solid orange lines represent the repulsive interaction between two substitutional extrinsic atoms (compare \fig{fig:energies-substitution}) referred to the most strongly bound configuration.
  }
  \label{fig:mixing-energies}
\end{figure}

The calculations reveal strongly bound configurations for all three elements, with $E_{2[X]}^b$ as high as $-2.4$, $-2.7$, and $-3.2\,\eV$ for V, Ti, and Re, respectively [\fig{fig:mixing-energies}]. In addition, as mentioned above, the individual mixed interstitial defects have a large and very anisotropic strain field that could result in defect reorientation. This anisotropy should manifest itself in the formation volume of the di-interstitial configuration $V^f(2[X-\W]_\W)$ relative to the individual defects $V^f([X-\W]_\W)$, which can be defined entirely analogously to the binding energy above as
\begin{align}
    \Delta V_{2[X]}^f = V^f(2[X-\W]_\W) - 2 V^f([X-\W]_\W).
    \label{eq:formation-volume}
\end{align}
Here, $V^f(2[X-\W]_\W)$ and $V^f([X-\W]_\W)$ denotes the formation volumes of the di-interstitial and (single) interstitial, respectively, which were computed as described in Refs.~\onlinecite{CenSadGil05, JedLinBen15, GhaErh15}.

\begin{figure*}
  \centering
  \includegraphics[scale=0.375]{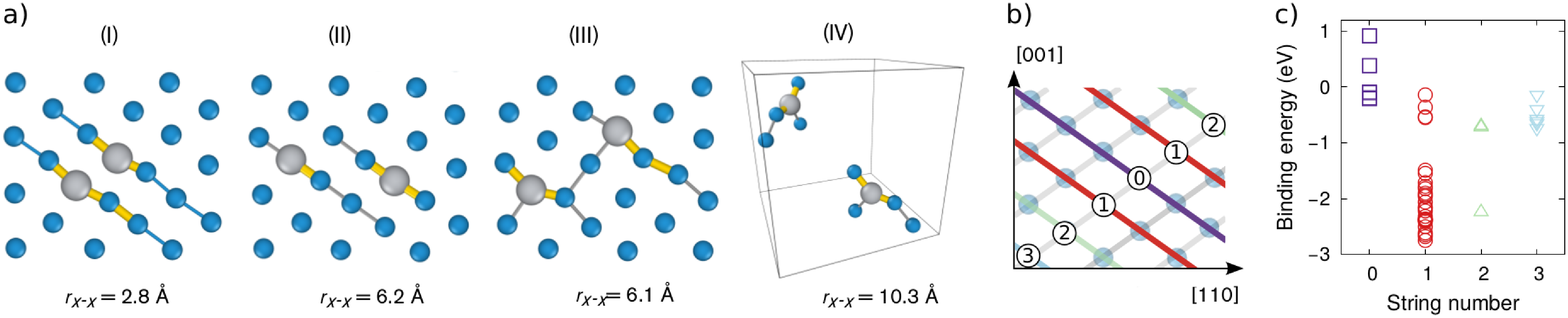}
  \caption{
    (a) Defect configurations involving two $(X-\W)_\W$ mixed--interstitials in which $X$ corresponds to V, Ti and Re atoms. The figure shows a slice parallel to a $\left\{110\right\}$ plane of the structure. Small (blue) spheres indicate tungsten atoms whereas large (gray) spheres indicate $X$ atoms. Thicker (yellow) cylinders indicate bond lengths shorter than 2.3\,\AA\ whereas thinner (gray) cylinders indicate bond lengths shorter than 2.5\,\AA.
    (b) An illustration of parallel $\left<111\right>$ strings in BCC tungsten.
    (c) Binding energy of a pair of titanium bridge mixed--interstitial with respect to string number.
  }
  \label{fig:double_interstitials}
\end{figure*}


In fact, we find a strong correlation between the binding energy $E_{2[X]}^b$ and the change in the formation volume $\Delta V_{2[X]}^f$ [\fig{fig:mixing-energies}].
While a strong binding is obtained for short defect separations, the lowest binding energies occur for somewhat longer distances. Yet, at the same approximate distance one can obtain binding energies that cover the entire range between strong and no binding [compare e.g., the data points near 6\,\AA\ in \fig{fig:mixing-energies}(a)]. This suggests that the relative orientation of defects is important.

A closer inspection of di-interstitial configurations reveals that in the structures that exhibit the strongest binding [structure (I) and (II) in \fig{fig:double_interstitials}(a)] the mixed-interstitials are aligned along parallel $\left<111\right>$ directions and the two strings hosting the extrinsic atoms are first nearest neighbors of each other [\fig{fig:double_interstitials}(b,c)]. If the mixed-interstitials are located either in the same string or in two strings that are second or farther nearest neighbors of each other [configuration (III) in \fig{fig:double_interstitials}(a)], the energy increases relative to the respective first-nearest neighbor configuration [\fig{fig:double_interstitials}(c)]. The behavior of defects that are located in the same string can be rationalized as being the result of strong repulsion due to overlapping strain fields. The situation for second and farther nearest neighbor configurations, on the other hand, is the result of a strong decrease in electronic coupling with distance. The short-range interaction between $\left<111\right>$ strings in BCC metals in general and W in particular is in fact also evident from the success of first-nearest neighbor string (Frenkel-Kontorova) models that have been used in the past to describe both interstitial and dislocation related features in these systems \cite{FitNgu08, ChiGilDud09, GilDud10, LiWurMot12}.
Finally, note that if the two mixed-interstitials are ``far'' apart and have no specific orientation relative to each other [configuration (IV) in \fig{fig:double_interstitials}(a)] their interaction vanishes and the binding energy decays to zero.

\begin{figure}
  \centering
  \includegraphics[scale=\myscale]{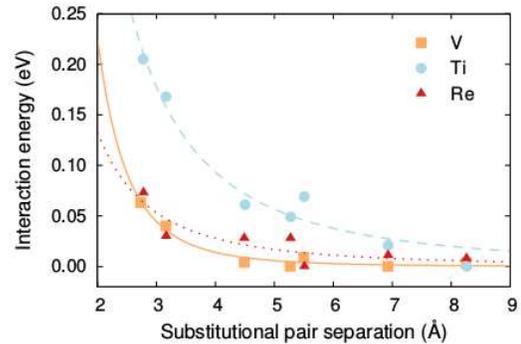}
  \caption{
    Energy of a pair of V, Ti and Re substitutional defects as a function of distance with respect to the infinite separation limit. Positive values indicate repulsion.
  }
  \label{fig:energies-substitution}
\end{figure}

It is worth noting that in the case of first-nearest neighbor strings, the binding energy decreases with increasing distance. 
This effect can be attributed to the repulsive interaction between two extrinsic atoms of the same type, as it exhibits the same strength and distance dependence as the interaction between two corresponding {\em substitutional} atoms in a BCC W matrix [compare Figs.~\ref{fig:mixing-energies} and \ref{fig:energies-substitution}].

\begin{figure*}
  \centering
  \includegraphics[scale=0.7]{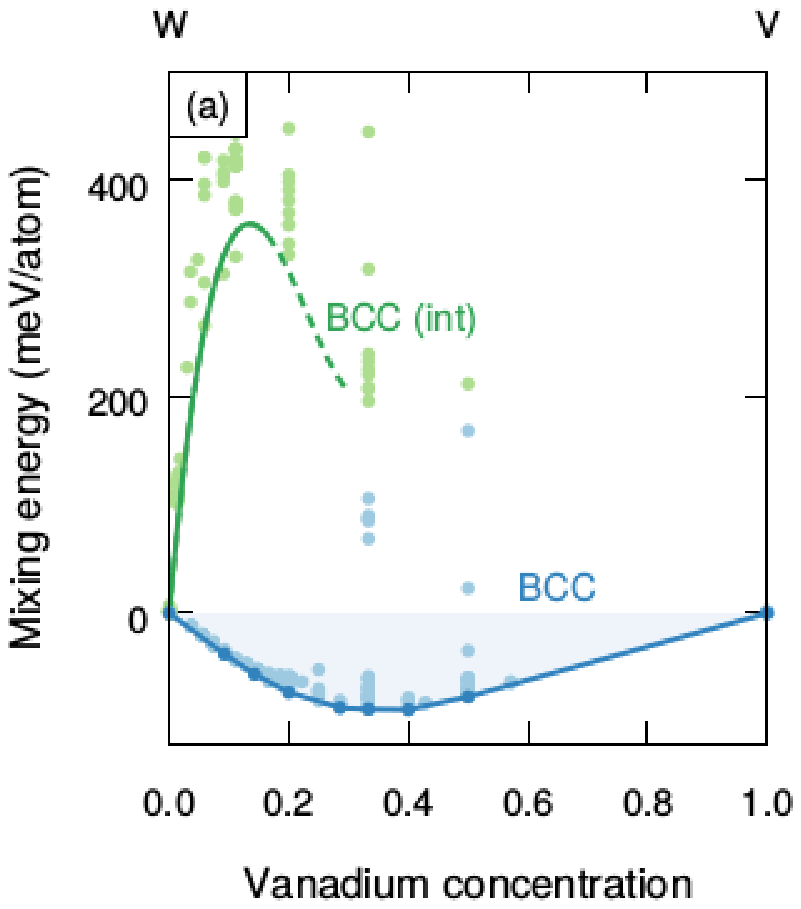}
  \includegraphics[scale=0.7,trim=30 0 0 0]{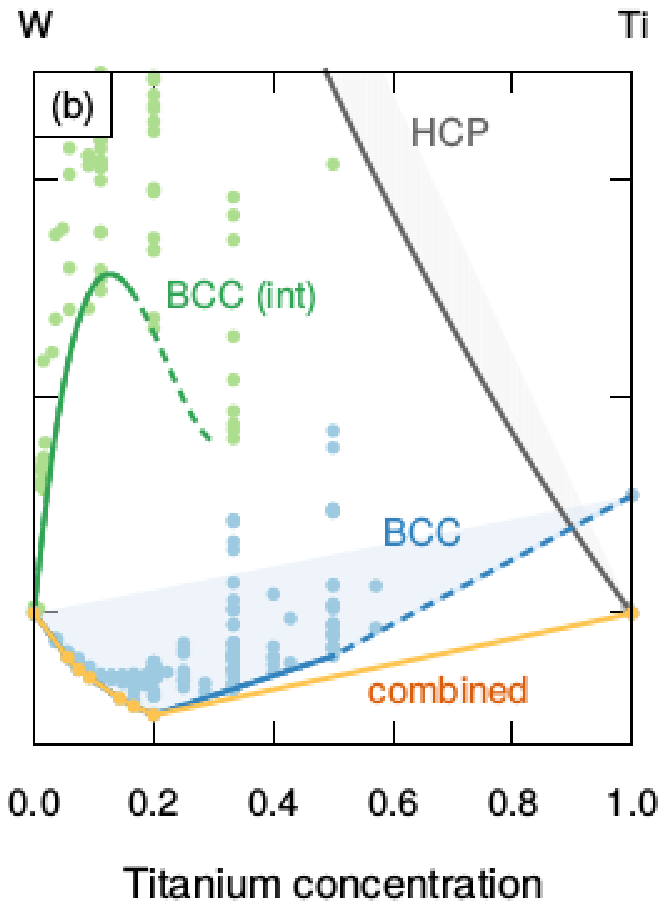}
  \includegraphics[scale=0.7,trim=30 0 0 0]{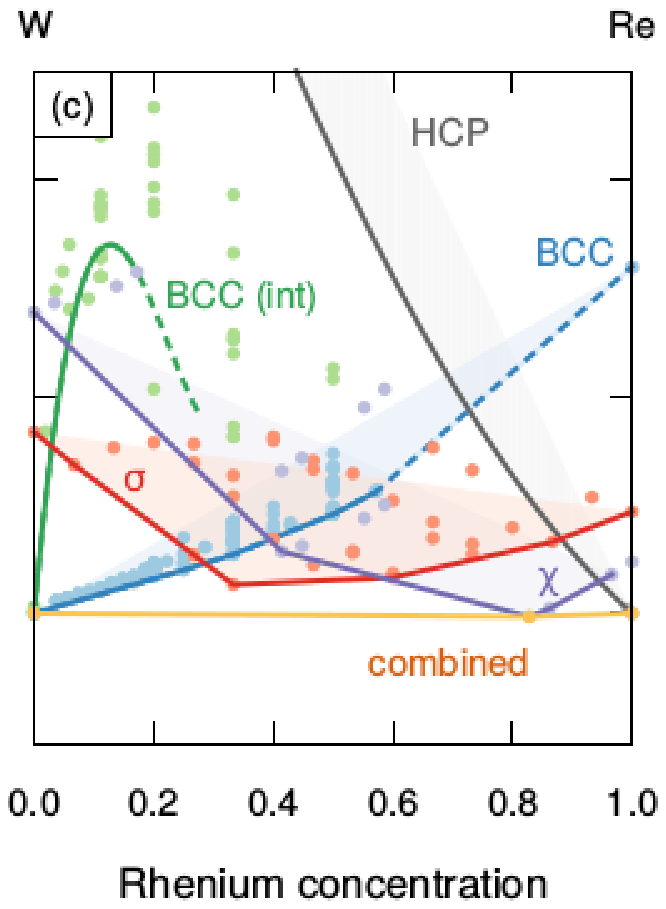}
  \caption{
    Mixing energy of (a) V, (b) Ti and (c) Re defects in tungsten-based structures obtained by substitution (squares) and interstitial insertion (circles). The solid (blue) line indicates the convex hull corresponding to the minimum mixing energy. The solid (yellow) line indicates the effective convex hull. Available data from W-V evaluated using cluster expansion are shown in plot (a) for comparison\cite{MuzNguKur11}. $\sigma$ and $\chi$ phases of a W-Re alloy\cite{CriJou10} are presented as well in plot (c) as a function of defect concentration
  }
  \label{fig:mixing_energy_BCC}
\end{figure*}

\subsection{Mixing energies and phase equilibria}
\label{sect:results-mixing-energies}

The results described in the previous section provide clear evidence of the energetic preference for the alignment of mixed-interstitials. The magnitude of the binding energies, however, suggests an effect on the electronic structure that extends beyond strain effects. To shed more light on this aspect, we now consider substitutional alloy configurations along with a series of interstitial based structures that were constructed as described in \sect{sect:method-structure-generation}. The mixing energy $E_{mix}$ for a given alloy composition $X_{x}\W_{1-x}$ was computed using the fully relaxed configurations according to
\begin{align}
  &E_{mix}(X_{x}\W_{1-x}) \nonumber\\
  &\quad=	E(X_{x}\W_{1-x}) - [ x E(X) + (1-x) E(\W) ],
\end{align}
where $x$ denotes the fraction of the alloying element $X$. Below we first summarize the results for the substitutional alloys before addressing the interstitial-based structures, which exhibit a rather universal behavior.

\paragraph*{Vanadium.}
Among the alloying elements included in this study V is the only other element with a ground state BCC lattice, and it therefore gives rise to the simplest mixing energetics [see \fig{fig:mixing_energy_BCC}(a)]. The difference in lattice constant at ambient conditions between W and V is 5.9\% (V: 2.989\,\AA, W: 3.177\,\AA\ according to PBE). The mixing energy is only slightly asymmetric and negative throughout the entire composition range. The low energy structures are closely spaced along the convex hull and our calculations closely match the data from Muzyk \textit{et al.} \cite{MuzNguKur11}

\paragraph*{Titanium.}
While at ambient conditions Ti adopts a hexagonal close-packed (HCP) structure ($\alpha$-phase), it also exhibits a BCC polymorph ($\beta$-phase), which occurs at temperatures above 1200\,K. Here we focus on BCC-based structures on the W-rich side of the composition range.

As in the case of W--V the mixing energy for BCC W--Ti is negative over the entire concentration range when referred to the elemental BCC phases [\fig{fig:mixing_energy_BCC}(b)]. Due to the energy difference between BCC Ti and HCP Ti, the mixing energy of BCC-based structure relative to the elemental ground states, i.e. BCC W and HCP Ti, however, turns positive at approximately 60\%. The lowest mixing energy occurs at a composition of 20\%. It belongs to space group R$\bar{3}m$ and has the same structure of the W$_4$V structure described in Ref.~\onlinecite{MuzNguKur11}.

W and Ti also mix on the HCP lattice as the energy of one substitutional W atom in HCP Ti is negative. The energy offset between HCP W and BCC W is, however, even larger than in the case of Ti and thus the mixing energy referred to the ground state structures is positive over the entire composition range. When combined, these data imply that one can expect a very asymmetric phase diagram with a large solubility for Ti in BCC-W but a very small solubility for W in HCP-Ti. In the high-temperature region of the phase diagram, which is experimentally accessible \cite{RudWin68}, this is indeed the case. For lack of reliable low temperature data, previous thermodynamic models of the phase diagram, however, assumed a mixing energy that is positive over the entire composition range \cite{KauNes75, LeeLee86}. As a result, these models predict the solubility in both limits as the temperature approaches zero. A revision of the low temperature part of the phase diagram is therefore in order, which will be the subject of a separate study \cite{GhaAngErh16}.

\paragraph*{Rhenium.}
As for Ti, the mixing energy of Re in W on BCC and HCP lattices is negative when referred to BCC and HCP elemental reference states, respectively (\fig{fig:mixing_energy_BCC}(c)). The HCP-BCC energy difference for Re is, however, even larger than for Ti, which causes the mixing energy for these two phases to be positive. This implies very small finite-temperature solubilities for both Re in BCC-W and W in HCP-Re in agreement with experiment.

In addition to the BCC and HCP phases, the W--Re system also features $\sigma$ and $\chi$ phases \cite{CriJou10}. The $\sigma$ phase [red line in \fig{fig:mixing_energy_BCC}(c); data from Ref.~\onlinecite{CriJou10}] experimentally occurs at high temperatures only, consistent with slightly positive mixing energies. The $\chi$ phase on the other hand [purple line in \fig{fig:mixing_energy_BCC}(c)] has a small negative mixing energy at about 83\%, which is compatible with this phase being a ground state (zero K) phase.

\paragraph*{Interstitial-based structures.}
Thus far it has been shown that while the crystal structures of the elemental boundary phases differ between V, Ti, and Re, the substitutional alloys show common characteristics with both BCC and HCP based phases being miscible with respect to the elemental phase with the same crystal structure. One also observes strong similarities with respect to the interstitial-based structures. While some of them relax into substitutional structures, the majority remain close to the initial structure after relaxation. For all three elements (V, Ti, Re) the mixing energies of these interstitial-based structures exhibit a similar dependence on concentration (green data points in \fig{fig:mixing_energy_BCC}). The mixing energies are positive and feature a maximum at about 15\%. For concentrations of approximately 30\%\ and above the mixing energy approaches zero and the structures loose interstitial-characteristics, which is apparent from an analysis of the pair distribution functions. This suggests that if the concentration of mixed-interstitials reaches a concentration of $\gtrsim\,30\%$ the structure becomes (locally) unstable with respect to a substitutional phase.

The concentration dependence of the mixing energy of interstitial-based structures thus suggests a miscibility-gap like behavior between dilute interstitial solutions and more concentrated structures, which contain 30\%\ or more of the alloying element and can transition more readily into substitutional configurations. These results further support the notion that interstitial agglomeration is energetically favorable.

\subsection{Mixed-interstitial migration}
\label{sect:results-migration}

\begin{figure}
   \centering
   \includegraphics[scale=0.09]{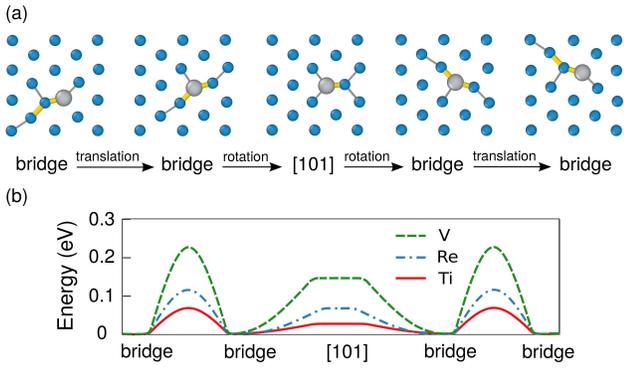}
   \caption{
     Energetically favorable non-dissociative diffusion of a mixed--interstitial in dilute tungsten alloy. The configurations are presenting the [$\bar{1}$10] dumbbell mixed--interstitial translation along [100] direction including a 90$^{\circ}$ rotation.
The plot in the lower row illustrates the corresponding migration barriers in eV.  
For a better comparison, the path, which connects bridge to [101] dumbbell mixed--interstitial via a rotation, is illustrated in the beginning of the migration barrier plot.
   }
   \label{fig:migration}
\end{figure}

Up to this point, we have shown that there is an energetic driving force for further interstitial agglomeration and that mixed-interstitials prefer an alignment along parallel nearest-neighbor $\left<111\right>$ strings. Next we discuss migration energies of interstitial structures. For all elements considered here the binding energy between a SIA and a substitutional defect is large (V: $-1.8\,\eV$, Ti: $-0.6\,\eV$, Re: $-0.8\,\eV$) \cite{BecDom09, SuzYamHas14, GhaErh15}. Thus, if mixed-interstitial diffusion were to proceed via dissociation of the mixed-interstitial, defect migration would be very slow except at very high temperatures. It has, however, been shown for the case of Re that mixed-interstitial can migrate via a non-dissociative mechanism with a much lower barrier, which in the case of Re was computed to be 0.12 eV \cite{SuzYamHas14}.

This low barrier is obtained for a sequence of a short displacive transformation, which shifts the center of the mixed-interstitial by $a_0\sqrt{3}/2$ along $\left<111\right>$, a rotation of the mixed-interstitial orientation, and another short jump along $\left<111\right>$ [\fig{fig:migration}(a)]. As a result of these events the interstitial center moves by one lattice constant along $\left<100\right>$.

Note that during the rotation the configuration effectively passes through a $\left<110\right>$ dumbbell configuration, which happens to be practically identical with the saddle point of that segment. The rotational barrier therefore closely corresponds to the energy difference between the bridge and the $\left<110\right>$ mixed-dumbbell, at least for V, Ti, and Re.

We have computed the landscape for this mechanism for all three extrinsic elements, see \fig{fig:migration}(b). For the initial translation barrier, we obtain values of 0.22, 0.07, and 0.11 eV for V, Ti, and Re, respectively. The latter value agrees well with earlier calculations \cite{SuzYamHas14}. The barriers for the rotation  are 0.14, 0.02, and 0.07\,eV for V, Ti, and Re, respectively. Since the rotational barrier is always smaller than the barrier for the translation segment, the effective barrier for the entire path is set by the latter. While Ti has the lowest barrier, the values for V and Re mixed-interstitials are similar in magnitude, so that these defects should be mobile under reactor relevant conditions. Note that the order of the barriers correlates with the degree of anisotropy of the formation volume tensor of the bridge interstitials \cite{GhaErh15}.

\section{Discussion}
\label{sect:discussion}

One of the main motivations behind this work is to investigate the causes behind the sub-soluble precipitation of transmutation solutes (mainly Re) observed in W subjected to high-energy particle irradiation. We seek mechanisms that connect point defects to solute transport in binary systems below the solubility limit. While long term precipitation kinetics is governed by the onset of defect fluxes to sinks, which is well beyond the scope of electronic structure calculations, the underlying mechanisms responsible for solute-defect coupling can only be investigated at the atomic scale using high accuracy methods. Here we have employed DFT calculations to study the fundamental energetics of interstitial-solute complexes in three different binary alloys, W-Re, W-Ti, and W-V.
These three are the main implications of our findings:
\begin{enumerate}
\item For all binary systems considered, mixed interstitial configurations are energetically favored over SIA-solute complexes. This is consistent with previous work on the subject conducted on W alloys \cite{KonYouWu14, SuzYamHas14, GhaErh15}. 
\item These mixed interstitials all display a migration mechanism that involves low-energy ($\leq0.14$ eV) rotations in the bridge interstitial configuration, followed by short translations in the $\langle 110\rangle$ orientation. In contrast with the \emph{interstitialcy} mechanism, which propagates the lattice perturbation leaving the solute behind, this `associative' mechanism can effectively transport solutes long distances in three dimensions, rendering it an efficient mass transport mechanism (in fact, due to the comparatively low migration energy barriers, interstitial-mediated transport could potentially surpass that due to vacancies \cite{SuzYamHas14}). 
\item While the heat of solution of all substitutional solutes at 0 K is negative in the BCC W matrix, corresponding to compounds that form solid concentrated solutions, the formation energies of all mixed interstitials are positive. This of course results in a strong segregation driving force for mixed interstitials defects, which ultimately may result in solute precipitation by way of mixed-dumbbell clustering.    
\end{enumerate}


Another aspect worth considering related to the precipitation kinetics of sub-soluble transmutants in W is the acicular (elongated) shape of the resulting precipitates.  Precipitates are seen to form in the intermetallic $\sigma$ and $\chi$ phases \cite{TanHasHe09, HasTanNog11, XuBecArm15}, although there is  experimental indication that they nucleate as solute-rich BCC clusters \cite{Edmondson15} that may ultimately undergo a structural transformation \cite{ErhMarSad13}. While the shape of these intermetallic precipitates is likely to be controlled by anisotropic interfacial energies and other thermodynamic features, here we put forward a mechanism based on the computed DFT energetics that may contribute to the incipient alignment of these clusters along crystallographic BCC $\langle111\rangle$ directions.
The mechanism is as follows:
\begin{enumerate}
\item[(i)]
  Starting from a substitutional alloy of W and Re, irradiation leads to the production of SIAs, which diffuse along $\langle111\rangle$ directions throughout the lattice until they become trapped by substitutional Re atoms due to a high binding energy.
\item[(ii)]
  Mixed interstitials migrate associatively in 3D eventually resulting in clustering. The strong energetic driving force for clustering (see \sect{sect:results-interstitial-binding}) and the low barriers for non-dissociative interstitial migration (see \sect{sect:results-migration}) facilitate defect co-alignment.
\item[(iii)]
  These mixed di-interstitials are kinetically very stable due to the strong binding energies, which makes them preferential sites for more defect absorption. While additional interstitials approaching these small clusters may not be initially aligned, they are energetically favored to again rotate into alignment, increasing cluster size and forming a precipitate nucleus.
\item[(iv)]
  As interstitial accumulation proceeds, the local Re concentration increases and the system samples the mixing energy curve representing interstitial-based configurations [see \fig{fig:mixing_energy_BCC}(c)]. 
\end{enumerate}

\begin{figure}
  \centering
  \includegraphics[scale=0.11]{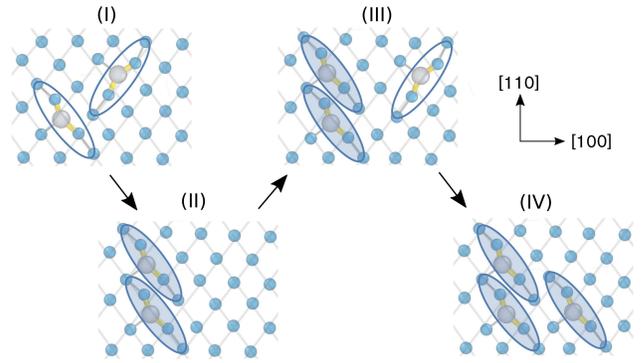}
  \caption{
    Schematic illustration of the path to elongated solute-rich clusters. Shown is an atomic slice parallel to a $\left\{110\right\}$ plane of the BCC lattice.
    (i) Initially mixed interstitials are trapped at substitutional extrinsic atoms with no particular orientation relationship.
    (ii) Via translations and rotations the defects move into alignment with respect to each other, preferably in neighboring parallel $\left<111\right>$ strings.
    (iii) Other interstitials are attracted to the nucleus moving relatively freely as long as they have not achieved alignment with the other defects in the cluster.
    (iv) Full alignment yields the lowest energy configurations.
  }
  \label{fig:schematic-mechanism}
\end{figure}

It is also worth mentioning that, as pointed out by Tanno and Hasegawa \cite{TanHasHe09, HasTanNog11}, the formation of precipitates occurs at high accumulated doses ($>5$ dpa), after sufficient transmutant buildup has occurred and the void-lattice stage subsides by recombination with large cascade-produced SIA clusters,  resulting in an excess population of single SIAs and/or mixed interstitials. 
The above mechanism for interstitial-mediated solute transport and clustering is schematically depicted in \fig{fig:schematic-mechanism}.
We recognize that this picture is somewhat speculative at this point, and we are currently implementing the energetics and the mechanisms presented here into long-term kinetic models capable of approaching doses of several dpa.  In any case, the formation of these precipitates is one of the possible pitfalls associated with the development of W-based alloys for applications in fusion environments, as $\sigma$/$\chi$ precipitates are known to cause very strong matrix hardening \cite{Snead2016}.

\section{Conclusions}

In conclusion, in the present study we have demonstrated that mixed interstitials in BCC-W involving V, Ti, and Re are strongly attracted to each other with binding energies of several eV. This interaction leads to the alignment of interstitials along parallel first nearest neighbor $\left<111\right>$ strings.

All three systems exhibit mixing on the BCC lattice with respect to BCC boundary phases. In the case of Ti and Re we also find negative formation energies for substitutional defects on the HCP lattice. The energetic offset between HCP-W and BCC-W on one side and BCC-Re and HCP-Re on the other side of the composition range, however, leads to very small solubilities in the boundary phases. In the case of Ti the BCC/HCP energy difference is relatively smaller, which gives rise to an asymmetric solubility that is substantially larger on the W-rich side.

Furthermore, we have calculated the activation barriers for mixed interstitial motion, which are comparably low for all three elements. The migration mechanism involves defect migration and reorientation, resulting in effective solute transport in 3D. 

The ease of rotation and strong binding of mixed interstitials could help explain the incipient formation of BCC solute-rich clusters oriented preferentially along $\langle111\rangle$ directions. It is not clear whether this may help steer the formation of elongated $\sigma$ and/or $\chi$ phase precipitates, as this is a more complex process involving long term kinetics and structural transformations, but it is interesting to note that there is an intrinsic directionality associated with the formation of these clusters. The analogies among V, Ti, and Re with respect to defect and alloying properties suggest that similar mechanisms could also come into play in W-Ti and W-V alloys.

The impact of interstitial binding described here on cluster formation and subsequently mechanical properties deserves further study. This is motivated by the observation that interstitial clustering has been shown give rise to complex diffusion behavior in other BCC metals \cite{fu2005multiscale,EVANS1983180} and to impact swelling behavior \cite{osetsky2003one}. 

\section{Acknowledgments}

This work was supported the {\em Area of Advance -- Materials Science} at Chalmers, the Swedish Research Council in the form of a Young Researcher grant, and the European Research Council via a Marie Curie Career Integration Grant. Com\-puter time allocations by the Swedish National Infrastructure for Computing at NSC (Link\"oping) and C3SE (Gothenburg) are gratefully acknowledged. JM acknowledges support from US-DOE's Early Career Research Program.

\end{document}